\documentstyle[12pt]{article}
\begin{document}
%%%%%%%%%%%%%%%%%%%%

\title{Evolution of Close Neutron Star Binaries}
\author{Wataru Ogawaguchi$^1$ and Yasufumi Kojima$^2$ \\ 
\\ 
\small{$^1$ Department of Physics, Tokyo Metropolitan University}\\ 
\small{Hachioji, Tokyo, 192-03 JAPAN}\\ 
\\ 
\small{$^2$ Department of Physics, Hiroshima University}\\ 
\small{Higashi-Hiroshima, Hiroshima, 739 JAPAN}} 
\date{\today}

\maketitle

\begin{abstract}
\qquad 

We have calculated evolution of neutron star binaries towards
 the coalescence driven by gravitational radiation.
The hydrodynamical effects as well as the general relativistic
 effects are important in the final phase.
All corrections up to post$^{2.5}$-Newtonian order
 and the tidal effect are included in the orbital motion.
The star is approximated by a simple Newtonian stellar model
 called affine star model.
Stellar spins and angular momentum are assumed to be aligned.
We have showed how the internal stellar structure affects 
 the stellar deformation, variations of the spins, and 
 the orbital motion of the binary just before the contact.
The gravitational wave forms from the last a few revolutions
 significantly depend on the stellar structure.

\end{abstract}
\newpage

%%%%%%%%%%%%%%%%%%%%1
\section{Introduction}
\qquad 

Direct detection of gravitational waves will be possible
 in not so far future by the recent constructing laser
 interferometers such as LIGO \cite{Abr92}. 
One of the promising sources is coalescing compact star binary
 consisting of neutron stars or black holes. 
The coalescence rate of the binary is estimated as
 $ \sim 10^{-7} $yr$^{-1}$ Mpc$^{-3}$ \cite{Nar91,Phi91}. 
If we can observe up to a few hundred Mpc,
 we have a few events by a year. 
The detection of the gravitational wave is not only an important
 probe of general relativity, but also a new tool of measurement
 of Hubble constant and high density nuclear matter
 \cite{Sch86,Cut93}. 

Theoretically, it is important to understand the final phase of
 the binary coalescence, that is, how the binary evolves
 from the separation distance of a few hundred km
 to the contact or disruption of the binary. 
The frequency of the gravitational radiation at this phase
 corresponds to the observational frequency band,
 about $10 \sim 10^3$ Hz.
The preparation of the gravitational wave form as the template
 is necessary for the matched filter analysis. 
When the separation $L$ of the binary is much larger than the
 stellar radius $R$, each star can be regarded as a point mass. 
The study to get the matched filter is currently in progress using
 the higher order post-Newtonian approximation \cite{Lin90,BDI95}. 
In particular, the test particle limit is extensively studied
 by the perturbation of the black hole space-time.
 (See e.g., Ref. \cite{Tag94a,Tag94b}.) 
The rotational velocity gradually increases with shrinking of
 the orbital radius, and therefore higher post-Newtonian
 corrections become crucial. 
When $L$ is a few times $R$,
 hydrodynamical effects also become important. 
The only method to examine the final stage may be solving
 the Einstein equations numerically. 
Available memory and speed of the computers are limited at present,
 so that the simulations are carried out with the Newtonian
 or post-Newtonian approximation, or preliminary results
 are obtained in fully relativistic case. 
(See Ref.\cite{Zhu94,Nak94} and therein.) 

%%%%
\par
One of hydrodynamical effects before the contact of the binary is
 tidal force. 
Lai, Rasio \& Shapiro calculated the dynamical evolution of
 the binary of simplified stellar models under the Newtonian
 gravity \cite{LRS1,LRS2,LRS3}. 
They showed that the orbit significantly deviates from that of two
 point masses. 
The reason is that there is additional contribution
 from the stellar quadrupole moment, which causes marginally stable
 circular orbit at $L =(2 \sim 3)R$. 
When the separation becomes below the critical distance,
 the radial infall velocity significantly accelerates
 in addition to the radiation reaction force. 
%%%
However, since the tidal potential depends on $L^{-6}$,
 the effect is significant only for close binary, i.e.,
 higher post-Newtonian correction. 
Natural question is whether the contribution is still significant,
 when other post-Newtonian corrections are included. 
The effect of the spinning star, i.e.,
 spin-orbit and spin-spin interaction,
 may be important since the tidal torque changes the spin. 
They are respectively post$^{1.5}$-Newtonian and
 post$^{2}$-Newtonian order in the magnitude. 
The first and second post-Newtonian corrections concerning
 the orbital motion are also important,
 when the tidal interaction becomes large. 
In this paper, we take all these corrections and
 the radiation reaction force into consideration,
 and study how the binary evolves at $L<15R$. 
We adopt simplified stellar models called affine star models.
The model is equivalent to the second order virial equation. 
In this approximation, the fluid displacement is limited to
 a certain class of the motion, i.e.,
 uniform expansion, rotation and quadrupole oscillation. 
The dynamical degrees of freedom of a star are reduced to
 finite number, and the equations of motion are not partial
 differential equations, but ordinary differential equations. 
Using this approximation, we can easily simulate
 the final phase of the binary with various effects
 concerning spin and quadrupole moment. 

In \S 2, we present our model. 
We show the collection for the interactions
 up to the post$^2$-Newtonian order and
 the radiation reaction force of post$^{2.5}$-Newtonian order. 
The affine star model is also briefly summarized. 
We show the deformation of the star, the orbital motion,
 the gravitational wave form and the evolution of the spin
 near the final stage just before the disruption of
 the binary in \S 3. 
In \S 4, we discuss the implication of our results. 

%%%%%%%%%%%%%%%%%%%%2
\section {Numerical Methods}

%%%2.1
\subsection{Orbital motion}
\qquad 

In order to solve orbital motion of the binary,
 we use the Hamiltonian formalism \cite{Dam88}. 
Regarding the star as a point mass with $M_a$ ($a=1,2$),
 we will examine the relative orbital motion. 
The motion for the center of mass will not be considered here. 
The equations of motion can be described by the total mass $M_{T}$
 and reduced mass $\mu$. 
We choose the orbital plane as the equatorial
 plane of the polar coordinate ($r, \theta, \phi$). 
The Newtonian, the first and second post-Newtonian equations of
 motion are determined by the following Hamiltonians. 
\begin{eqnarray}
H_N & = & 
 \frac{1}{2 \mu} \left( p_r^2 + \frac{p_{\phi}^2}{r^2} \right) 
 - \frac{G \mu M_T}{r}, 
\\
H_{PN} & = &
 \frac{3\nu-1}{8 c^2 \mu^3} \left( p_r^2 + \frac{p_{\phi}^2}{r^2}
 \right)^2
 - \frac{G M_T}{2 c^2 \mu r} \left\{ \left( 3 + \nu \right)
 \left( p_r^2 + \frac{p_{\phi}^2}{r^2} \right) + \nu p_r^2 \right\}
 + \frac{G^2 \mu M_T^2}{2 c^2 r^2}, 
\\
H_{P^2N} & = & 
 \frac{1 - 5 \nu + 5 \nu^2}{16 c^4 \mu^5}
 \left( p_r^2 + \frac{p_{\phi}^2}{r^2} \right)^3
 \nonumber\\ 
& &
 + \frac{G M_T}{8 c^4 \mu^3 r} \left\{ \left( 5 - 20 \nu - 3 \nu^2
 \right)
 \left( p_r^2 + \frac{p_{\phi}^2}{r^2} \right)^2
 - 2 \nu^2 p_r^2 \left( p_r^2 + \frac{p_{\phi}^2}{r^2} \right)
 - 3 \nu^2 p_r^4 \right\}
 \\
& &
 + \frac{G^2 M_T^2}{2 c^4 \mu r^2}
 \left\{ \left( 5 + 8 \nu \right) 
 \left( p_r^2 + \frac{p_{\phi}^2}{r^2} \right) + 3 \nu p_r^2
 \right\}
 - \frac{G^3 \left( 1 + 3 \nu \right) \mu M_T^3}{4c^4 r^3}, 
 \nonumber
\end{eqnarray}
 where $\nu = \mu/M_T$. 
\par

%%%%
Next, we consider the effect of the quadrupole moment of a star. 
If the star has mass quadrupole moment,
 the Newtonian gravitational potential has
 an additional contribution, $\sim 1/r^3$. 
The relative motion is affected by the monopole-quadrupole
 interaction \cite{BHMP}, 
\begin{equation}
H_T = - \frac{G}{2r^3} \left( 3n^i n^j - \delta^{ij} \right)
 \left\{ M_2 \left( {\bf{I}}_1 \right)_{ij}
 + M_1 \left( {\bf{I}}_2 \right)_{ij} \right\}, 
\label{eqtf}
\end{equation} 
 where ${\bf{n}} = {\bf{r}}/r$ and
${\bf{I}}_a$ is mass quadrupole moment of star $a$ ($a=1,2$). 

The spin-orbit and spin-spin interaction term
 in appropriate coordinate system can be written as \cite{BHMP}, 
\begin{equation}
H_{SO} = \frac{G}{c^2 r^3} {\bf{L}}_N \cdot
 \left\{ \left( 2 + \frac{3M_2}{2M_1} \right) {\bf{J}}_1
 + \left( 2 + \frac{3M_1}{2M_2} \right) {\bf{J}}_2 \right\}, 
\label{eqso}
\end{equation}
\begin{equation}
H_{SS} = \frac{G}{c^2 r^3}
 \left\{ 3\left( {\bf{J}}_1 \cdot {\bf{n}} \right)
 \left( {\bf{J}}_2 \cdot {\bf{n}} \right)
 - \left( {\bf{J}}_1 \cdot {\bf{J}}_2 \right) \right\}, 
\label{eqss}
\end{equation}
 where
 ${\bf{J}}_a$ is spin angular momentum of star $a$ and
 ${\bf{L}}_N$ is orbital angular momentum of the Newtonian order, 
 i.e., ${\bf{L}}_N = {\bf{r}} \times {\bf{p}}$. 
\par

Now, we shall estimate the order of magnitude for
 these various forces here. 
Suppose that a binary system consists of compact stars with
 mass $M_a \sim M$ and radius $R_a \sim R$ ($a=1,2$).
The orbital velocity $v$ at the separation distance $L$
 is $v \sim (G M/L)^{1/2}$. 
We define the dimensionless relativistic factor, $\epsilon$ as
 $\epsilon = (G M/ L c^2) \sim (v/c)^2$. 
Since the radius of the compact star is of order $G M/c^2$,
 the spin angular momentum $J$ and the quadrupole moment $I$
 can be written as $J = \alpha G M^2/c$ and $I = \beta G^2 M^3/c^4$,
 where $\alpha$ and $\beta$ are dimensionless numbers. 
The maximumly rotating Kerr black hole corresponds to $\alpha =1$. 
It is likely that $\alpha <1$. 
The value of $\beta$ is small in isolated system, but is
 induced by the companion star as $\beta \propto$ 
 $(R/L)^3 \sim \epsilon^3$. 
\par

We shall write the Newtonian force as
 $ f_N = G M^2 / L^2 $, which can be estimated as
 $ \partial H_N /\partial r \sim H_N /L $. 
The order of the first and second post-Newtonian corrections are
 $ \epsilon f_N $ and $ \epsilon^2 f_N $, respectively. 
The tidal interaction is the Newtonian force, but
 the magnitude for the compact star is
 $ G I M /L^4 $ $ \sim \beta \epsilon^2 f_N $
 $ \sim \epsilon^5 f_N $. 
This is higher post-Newtonian correction. 
The spin-orbit coupling term is formally the post-Newtonian
 correction, since the term contains the factor $G/c^2$. 
The magnitude is, however, estimated as
 $G/c^2 \times MLv \times J/L^4 \sim \alpha \epsilon^{3/2} f_N$,
 that is, the spin-orbit force is the post$^{1.5}$-Newtonian
 correction practically. 
In the similar way, the spin-spin interaction is also formally
 the post-Newtonian correction, but practically
 the post$^{2}$-Newtonian correction as
 $G/c^2 \times J^2 /L^4 \sim \alpha^2 \epsilon^2 f_N$. 
The equations of motion can be determined by the sum of
 these Hamiltonians.

%%%2.2
\subsection{ Affine star model}
\qquad 

As seen in the previous section, the orbital motion is affected
 by the spin angular momentum $J_i$ and the mass quadrupole moment
 $I_{ij}$, as well as the mass $M$. 
The spin and the quadrupole moment are determined
 by the stellar structure. 
We adopted so-called the affine star model for the stellar model. 
In this section, we will summarize the single affine star model
 \cite{CL85,LC86}. 
The binary consists of two affine star models ($a=1,2$). 
In this section, we omit the suffix $a$ and use
 the Cartesian coordinate system, whose origin is
 the mass center of a single star. 

\par
The position of a fluid element inside the star,
 $x_i(t)$ is specified by the Lagrangian map
 from its initial position ${\hat x}_i$ as 
\begin{eqnarray} 
x_i(t) = q_{ia}(t){\hat x}_a, 
\end{eqnarray}
 where $q_{ia}$ is $3\times 3$ matrix and $q_{ia}(0) =\delta_{ia}$. 
The initial sphere of radius $R$ is
 transformed into an ellipsoid as 
\begin{eqnarray} 
S_{ij}^{-1} x_i x_j = {R}^2,
\end{eqnarray}
 where $S_{ij}^{-1}$ is the inverse of a symmetric matrix,
 $S_{ij}=q_{ia} q_{ja}$. 
The eigenvalues of $S_{ij}$ determine three principal axes
 of the ellipsoid. 
\par

The dynamics of the fluid motion can be determined
 in the Hamiltonian formalism by the generalized coordinate $q_{ia}$
 and its conjugate momentum $p_{ia}=I_0 {\dot q}_{ia}$,
 where the inertial moment $I_0$ is described
 $I_0={1 \over 3} \int \hat{x}_a \hat{x}_a dM$
 and the dot denotes time derivative. 
The initial state is a spherically symmetric star with mass $M$
 and radius $R$, and we assume a polytropic equation of state. 
So that we have $ I_0 = {1 \over 5} \kappa_n M {R}^2$,
 where $\kappa _n$ is a constant relating with polytropic index $n$. 
The Hamiltonian concerning the internal fluid motion is
 the sum of the kinetic energy $T$, internal energy $U$,
 and self-gravitational energy $\Phi$, 
\begin{eqnarray} 
H_I = T + U + \Phi. 
\end{eqnarray}
The kinetic energy is a function of $p_{ia}$ given by 
\begin{eqnarray} 
T = {1 \over 2} \int {v_i}{v_i}dM = {1 \over {2 I_0}} p_{ia}p_{ia}. 
\end{eqnarray}
The internal energy is a function of $q_{ia}$. 
We use a polytropic equation of state, $P = \kappa \rho^{1+(1/n)}$,
 then the internal energy of the star is 
\begin{eqnarray} 
U = \int {n P \over \rho} dM = U_0 \left| {\bf q} \right|^{-1/n},
\end{eqnarray}
 where $\left| {\bf q} \right|$ is determinant of $q_{ia}$,
 and $U_0$ is the initial internal energy. 
The self-gravitational energy of the star can be calculated
 by an elliptical integral \cite{Cha69} 
\begin{eqnarray} 
\Phi = {1 \over 2} \Phi_0 S_{ik}
 \int_0^\infty { \left( {\bf S} + u{\bf \delta} \right)_{ik}^{-1} 
 \over { \left| {\bf S} + u {\bf \delta} \right|^{1 \over 2}} } du, 
\end{eqnarray} 
 where $\Phi_0$ is the self-gravitational energy
 of spherically symmetric star. 
Since the pressure force and self-gravity assumed to balance
 at the initial state, we have 
\begin{eqnarray} 
U_0 = - {\Phi_0 \over 3} = \frac{G M^2}{\left( 5-n \right) R}. 
\end{eqnarray} 
The quadrupole moment and the spin of the affine star model
 are given by 
\begin{eqnarray} 
I_{ij}=I_0 q_{ia} q_{ja}, 
\end{eqnarray} 
\begin{eqnarray} 
J_i = \varepsilon_{ijk} q_{ja} p_{ka}. 
\label{defspn}
\end{eqnarray} 
%

%%%2.3
\subsection{Equations of motion and gravitational radiation}
\qquad 

The dynamical degrees of freedom of the system are reduced
 to finite number in our approximation. 
The variables are fluid ones ($q_{ij}, p_{ij}$)
 and orbital ones ($r, \phi, p_r, p_\phi $). 
The system is determined by the following total Hamiltonian, 
\begin{equation}
H = H_{I1}+H_{I2}+H_{N}+H_{PN}+H_{P^2N}+H_{T}+H_{SO}+H_{SS}. 
\label{eqtotal}
\end{equation}
The evolution of ${\bf J}_a$ ($a=1,2$) is governed
 by the familiar formula \cite{BHMP}, 
\begin{eqnarray}
{ d{\bf J}_a \over{dt} } & = & {G \over c^2 r^3}
 \left\{ \left( 2+{3 \over 2} {M_b \over M_a} \right) {\bf L}_N
 - {\bf J}_b
 + 3 \left( {\bf n} \cdot {\bf J}_b \right) {\bf n} \right\}
 \times {\bf J}_a
\nonumber \\ 
 & & - { 3 G M_b \over r^3 }
 \left( {\bf n} \times \left( {\bf I}_a : {\bf n} \right) \right),
\label{eqprec}
\end{eqnarray}
 where the subscript $b$ means that $b=2$ for $a=1$,
 and $b=1$ for $a=2$. 
The first term of eq.($\ref{eqprec}$) represents
 the precession due to the spin-orbit and spin-spin coupling. 
This term changes the direction, but not the magnitude of the spin. 
Therefore, when the spins and angular momentum are aligned
 in the same direction, the first term vanishes and
 the direction of the spins is fixed. 
The second term shows that the magnitude of the spin changes
 through the tidal torque. 
It is easily checked that the above equation of spin is
 automatically satisfied, if ($q_{ij}, p_{ij}$) follow
 the equations of motion. 
Therefore, we do not have to calculate the evolution of spins
 additionally. 
\par

The system ($\ref{eqtotal}$) is conserved one. 
We add the effect of the gravitational radiation by
 the lowest quadrupole formula,
 which is post$^{2.5}$-Newtonian order. 
The dissipative force is introduced by adding
 the radiation reaction forces ($f_r, f_\phi, F_r, F_\phi$)
 in the Hamiltonian formalism (e.g., Ref.\cite{RRF,KoSc}). 
The explicit radiation reaction forms are given by 
\begin{eqnarray}
f_r & = & -\frac{8G^2}{15 c^5 \nu r^2}
 \left( 2p_r^2 + \frac{6p_\phi^2}{r^2} \right), 
\\
f_\phi & = & -{8 G^2 \over 3 c^5} \frac{p_r p_\phi}{\nu r^4}, 
\\
F_r & = & -{8 G^2 \over 3 c^5} \frac{p_r}{r^4}
 \left( \frac{p_\phi^2}{\nu r} - {G \nu M^3 \over 5} \right), 
\\
F_\phi & = & -{8 G^2 \over 5 c^5} \frac{p_\phi}{\nu r^3}
 \left( \frac{2G \nu^2 M^3}{r} + \frac{2p_\phi^2}{r^2} - p_r^2
 \right). 
\end{eqnarray}
\par

The evolution of the variables is determined by the following set
 of equations, 
\begin{eqnarray}
\frac{dr}{dt} &=& \enskip \frac{\partial H}{\partial p_r} + f_r,
\\ 
\frac{d\phi}{dt} &=& \enskip \frac{\partial H}{\partial p_\phi} +
 f_\phi, 
\\
\frac{dp_r}{dt} &=& -\frac{\partial H}{\partial r} + F_r, 
\\
\frac{dp_\phi}{dt} &=& -\frac{\partial H}{\partial \phi } + F_\phi, 
\\
\frac{dq_{ij}}{dt} &=& \enskip \frac{\partial H}{\partial p_{ij}}, 
\\
\frac{dp_{ij}}{dt} &=& -\frac{\partial H}{\partial q_{ij}}. 
\end{eqnarray}
\par

We describe two polarization modes of gravitational waves
 by $h_+$ and $h_\times$. 
They are written by the distance to the source $D$ and
 the directional angle $\Theta$. 
Including the quadrupole moment terms of each ellipsoids,
 the polarization modes are given by, 
\begin{eqnarray}
h_+ & = & { G\over 2 c^4 D} \left( 1 + \cos^2 \Theta \right)
 \nonumber \\ & &
 { d^2 \over dt^2} \left[ \mu r^2 \cos 2 \phi
 + \sum_{a=1,2} \left\{ ( {\bf{I}}_a )_1 - ( {\bf{I}}_a )_2 \right\}
 \cos 2 \left( \phi + \theta_a \right) \right], 
\label{eqgw1} 
\\ 
h_\times & = & { G\over 2 c^4 D} \left( 2 \cos \Theta \right)
 \nonumber \\ & &
 { d^2 \over dt^2} \left[ \mu r^2 \sin 2 \phi
 + \sum_{a=1,2} \left\{ ( {\bf{I}}_a )_1 - ( {\bf{I}}_a )_2 \right\}
 \sin 2 \left( \phi + \theta_a \right) \right], 
\label{eqgw2}
\end{eqnarray} 
 where $( {\bf I}_a )_i$ ($i=1,2$) is the inertial moment of
 the ellipsoid along the principal axis $a_i$. 
Both $a_1$ and $a_2$ axes are chosen on the orbital plane
 of the binary. 
The $a_1$ axis is initially chosen as the separation direction
 ${\bf n}$ of the binary and the $a_2$ axis is orthogonal to it. 
Three eigen-values of the matrix $({\bf I}_a)_{ij}$ give
 $( {\bf I}_a )_i$ ($i=1,2,3$). 
In eqs.(\ref{eqgw1}), (\ref{eqgw2}), $\theta_a$ is the lag angle,
 which is defined by the angle between the major axis of
 the ellipsoid $a$ and the direction to the companion star. 
Tidally locked state corresponds to $\theta_a=0$.

%%%%%%%%%%%%%%%%%%%%3
\section{Results}
\qquad 

We have calculated the evolution of close binary
 from a circular orbit of the separation
 $r=15R$ to the contact, using two identical affine stars. 
The most natural orbit near $r=15R$ is circular one, because
 the eccentricity is diminished by the gravitational radiation
 as the binary orbit shrinks. 
As for the internal structure at $r=15R$, we adopt
 the equilibrium solution of the ellipsoid, i.e., irrotational
 Roche-Riemann ellipsoid or Maclaurin-Jacobi-Roche ellipsoid
 \cite{Koch92}. 
The former solution basically represents tidally locked,
 static figure in the inertial frame. 
We adopt Maclaurin-Jacobi-Roche solution
 as the spinning stellar model. 
Both stellar spins are assumed to be parallel to the orbital
 angular momentum, but the initial magnitude of spin is variable. 
The magnitude is parameterized by
 $\chi = \left| {\bf J}_a \right| / ({\bf I}_a)_3 \Omega_o$
 at the initial separation,
 where $\Omega_o$ is the orbital angular frequency of the star. 
We selected the parameter as $\chi = -2, -1, -0.5, 0, 0.5, 1, 2$. 
The maximum rotation $\left| \chi \right| =2$ in our model
 corresponds to the period of $ \sim 40 \pi t_D$,
 where $t_D$ is the dynamical time at an individual star surface
 i.e., $t_D = ( R^3/GM )^{1/2}$. 
Typically, it is about ten milliseconds for the actual neutron star. 
The results given below are almost for $\chi=0$ except
 in \S 3.3 and \S 3.4,
 because the results for $\chi \neq 0$ models are almost
 the same as those for $\chi=0$. 
The stellar model is set up as $ (GM/R c^2) \simeq 0.2 $
 with polytropic index $n=0.5$ or $n=1$,
 corresponding to the typical neutron star of
 $M \sim 1.4M_\odot$ and $R \sim 10$ km. 
We have numerically calculated the subsequent evolution
 from these initial conditions. 
We ignored the dissipative viscosity terms for internal motions
 during the evolution of the orbit, so that the stellar circulation
 is conserved before tidal torque is important. 
We will give our results of numerical simulations below.

%%%3.1
\subsection{Tidal distortion}
\qquad 

We show the typical examples of the quadrupole deformation
 of stars with $n=0.5$ and $n=1$ in Fig.1. 
The initial state corresponds to $\chi = 0$,
 but the results are almost the same for $\chi \neq 0$. 
Three orthogonal principal axes, $a_1$, $a_2$ and $a_3$,
 of the ellipsoid are shown as a function of
 the orbital separation. 
The minimum separation distance corresponds to
 the contact of the binary. 
That is, $r \sim 2.4R$ for $n=0.5$ and $r \sim 2.3R$ for $n=1$. 
The axes $a_1$ and $a_2$ are located on the orbital plane,
 while $a_3$ is the direction of the orbital angular momentum. 
The $a_1$-axis is initially chosen as the separation direction
 ${\bf n}$ of the binary, but the direction does not exactly follow
 the orbital motion during the dynamical evolution. 
The orbital revolution at small $r$ becomes so fast that
 the principal axis slightly deviates from the direction ${\bf n}$,
 i.e., there is a time lag. 
We will discuss the lag angle in next subsection. 
However, the lag angle is small, so that the star is elongated
 almost in the separation direction ${\bf n}$
 and compressed in other two directions. 
The quadrupole deformation becomes significant for
 $r < 5R$ and grows as $\left| a_i / R -1 \right| \propto r^{-3}$. 
The internal structure also affects the shape of the ellipsoids. 
The model with $n=1$ is more centrally condensed
 than that with $n=0.5$, so that the shape is less deformed. 
The difference is, however, a few percent in the magnitude. 
We will see how this difference appears in the orbital motion
 and gravitational radiation below.

%%%3.2
\subsection{Orbital motions}
\qquad 

The binary evolution driven by the gravitational radiation is
 calculated in terms of four approximations
 to estimate various forces. 
(1) All post$^{2}$-Newtonian, spin and tidal effects are included. 
(2) The post$^{2}$-Newtonian effects are included,
 but the stars are regarded as point masses. 
(3) Purely Newtonian gravity is used,
 but the tidal effects are included. 
(4) The Newtonian gravity is used,
 and the stars are regarded as point masses. 
We show the angular and radial velocities in Fig.2 and Fig.3. 
We give the comparison between the approximations (1) and (2)
 in Fig.2a and Fig.3a, and the comparison between
 the approximations (3) and (4) in Fig.2b and Fig.3b. 
The difference due to the approximations becomes
 clear below $r \sim 3 R$, where the star is much deformed. 
The model with $n=0.5$ is more deformed and therefore
 the deviation from the point mass approximation
 becomes more significant. 
Although the spin-orbit force with positive $\chi$ are repulsive,
 the post-Newtonian corrections, spin-spin and tidal force
 are dominant attractive forces. 
Combined with all these effects, the final approaching velocity 
 significantly accelerates. 

%%%
\par
The absolute magnitude of the velocity shifts,
 when the post-Newtonian corrections are included. 
The ratio, i.e., the direction of colliding stars,
 may be more useful than the absolute values
 as the initial conditions of the hydrodynamical simulation
 after the coalescence. 
We show the velocity ratio $-V_r / V_{\phi}$ in Fig.4. 
The approaching velocity increases up to ten percent
 of the orbital angular velocity, i.e.,
 the velocity ratio $-V_r /V_{\phi} \sim 0.1$
 at the contact of the binary. 
This ratio further increases with decrease of
 the polytropic index $n$, because the tidal force is larger. 

%%%%
\par
When the binary separation is large enough,
 the binary evolves at much longer time-scale
 compared with that of the internal motions. 
The stellar figure is tidally locked at this stage. 
As the time-scale of the orbital motion decreases,
 the tidal lock becomes loose and the lag angle becomes non-zero. 
We show the lag angles for two stellar models
 ($n=1 $ and $n=0.5$) in Fig.5. 
The negative lag angles means that the elliptical figure
 does not follow the orbital revolution, and that
 there is a time lag to settle down. 
The lag angles amount to 0.1 radian just before the contact. 
This suggests that both stars at the coalescence
 touch each other not in the longest axis of the ellipsoid,
 but the direction slightly deviates.

%%%3.3
\subsection{Gravitational wave forms}
\qquad 

The tidal force affects the orbital motion just before the contact. 
We have synchronized the phase of the gravitational wave
 at the initial separation $r=15R$. 
In Fig.6 and Fig.7, we only show the evolution of
 the gravitational wave forms radiated
 at the last several revolutions. 
The correction of tidal force in the amplitude is of order
 $I/ \left( \mu r^2 \right) \sim \beta \varepsilon^2$. 
Therefore, tidal effect is a small correction in the amplitude,
 but it accumulates in the phase of the wave. 
As the time proceeds, the wave form deviates from that of
 point masses, which is represented by the dotted line in Fig.6. 
The deviation of the $n=0.5$ model is larger than that of
 the $n=1$ model, because of larger tidal deformation. 
The phase shift becomes significant only for
 the last a few revolutions, and the amount of the phase shift
 is less than $\pi$, if there is no initial spin. 
In Fig.7, we show the dependence of the initial spin. 
The stellar spin gives large difference in the phase shift. 
For the larger initial spin, the repulsive spin-orbit force becomes
 stronger, so that it takes much longer time to contact each other. 
The amount of the phase lag becomes more than $\pi$
 for the fast spinning case. 

%%%
\subsection{Evolution of spin}
\qquad 

The stellar spin is modified through the tidal torque
 following eq.($\ref{eqprec}$). 
We have calculated the evolution of the spin
 from various initial states. 
We show the typical evolution of the spin in Fig.8. 
The variations of spins are quite small at large separation, 
 but monotonically grow only in the final stage, $r < 4R$. 
The stars significantly deviate from the spherical state
 as mentioned in \S 3.1, and the tidal torque
 becomes quite large there. 
Therefore, variation of the spin is quite a steep function
 of $r$, since the tidal torque is
 $dJ /dt \propto I r^{-3}$ $\propto r^{-6}$. 
We also show the total amount of the change of the spin
 for the various initial spin angular momentum in Fig.9. 
The transfered momentum through the tidal torque
 decreases with the increase of the parameter $\chi$. 
The angular momentum is much more transfered
 in the counter-rotating star with the orbital angular momentum,
 i.e, star with negative $\chi$.

%%%%%%%%%%%%%%%%%%%%4
\section {Discussion}
\qquad

The simple estimate of the tidal effect is of order
 $\sim \beta \epsilon^2 f_N \sim \epsilon^5 f_N$,
 as shown in \S 2.1. 
The final state of the binary corresponds to $\epsilon \sim 0.3$,
 so that the correction seems to be $(0.3)^5 \sim 0.002$,
 but is much large in the actual problem. 
The discrepancy comes from the existence of
 the marginal circular radius,
 where the hydrodynamical instability occurs \cite{LRS1}. 
If the radiation reaction force is not included,
 there are circular orbits for any separations,
 but they are not always stable \cite{KWW93}. 
We expect that there exists innermost stable circular orbit,
 which the post-Newtonian gravity and the tidal force cause
 at the critical separation. 
Two stars significantly accelerate in the radial direction
 below this critical radius. 
In a future paper, we will show that the location of
 the innermost stable circular orbit is affected
 by polytropic index $n$ and initial spin parameter $\chi$. 
The tidal effects at the last a few cycles cause
 large radial infall velocity, lag angle and spin-up. 
The higher post-Newtonian corrections or fully relativistic
 hydrodynamical calculation will be required to study
 much more correct final phase. 
The topic treated in this paper is preliminary,
 but will be useful as the guide for the future theoretical work
 and the observations by the advanced type
 of the laser interferometer.

%%%%%%%%%%%%%%%%%%%%
\section*{Acknowledgment}
\qquad 

This work was supported in part by the Japanese Grant-in-Aid for
 Scientific Research of the Ministry of Education,
 Science and Culture (No.05218208, 06740225). 

\newpage

%%%%%%%%%%%%%%%%%%%%

\newpage

%%%%%%%%%%%%%%%%%%%%
\section*{Figure Captions}

{\bf Fig.1a:} 
Variations of the tri-axes of ellipsoid with $n=0.5$ and $\chi=0$
 as a function of the binary separation. 
The solid, dashed, and dotted lines indicate
 the axis $a_1$, $a_2$, and $a_3$. 
\\
\\ 
{\bf Fig.1b:}
The same as Fig.1a, but with $n=1$. 
\\ 
\\
{\bf Fig.2a:} 
The angular velocity at the binary separation $r$. 
The solid line denotes the result included
 post$^{2}$-Newtonian, spin-orbit,
 spin-spin and tidal forces with $n=0.5$ stellar model
 without initial spin. 
The dashed line is the same but with $n=1$ model. 
For comparison, the result of point mass,
 neglecting the tidal and spin effects,
 is described by the dotted line. 
\\
\\ 
{\bf Fig.2b:} 
The same as Fig.2a, but the results are based on
 the purely Newtonian calculation. 
The solid and dashed line denote the results included
 tidal forces with $n=0.5$ and $n=1$ models, respectively. 
The dotted line is the result of point mass. 
\\ 
\\
{\bf Fig.3a:} 
The radial velocity of the binary with relativistic corrections. 
The meaning of each line is the same as in Fig.2a. 
\\
\\ 
{\bf Fig.3b:} 
The radial velocity of the binary with Newtonian gravity. 
The meaning of each line is the same as in Fig.2b. 
\\ 
\\
{\bf Fig.4a:}
The velocity ratio $-V_r/V_{\phi}$ in the coalescing binary
 with relativistic corrections. 
The meaning of each line is the same as in Fig.2a. 
\\
\\ 
{\bf Fig.4b:}
The velocity ratio $-V_r/V_{\phi}$ in the coalescing binary
 with Newtonian gravity. 
The meaning of each line is the same as in Fig.2b. 
\\
\\ 
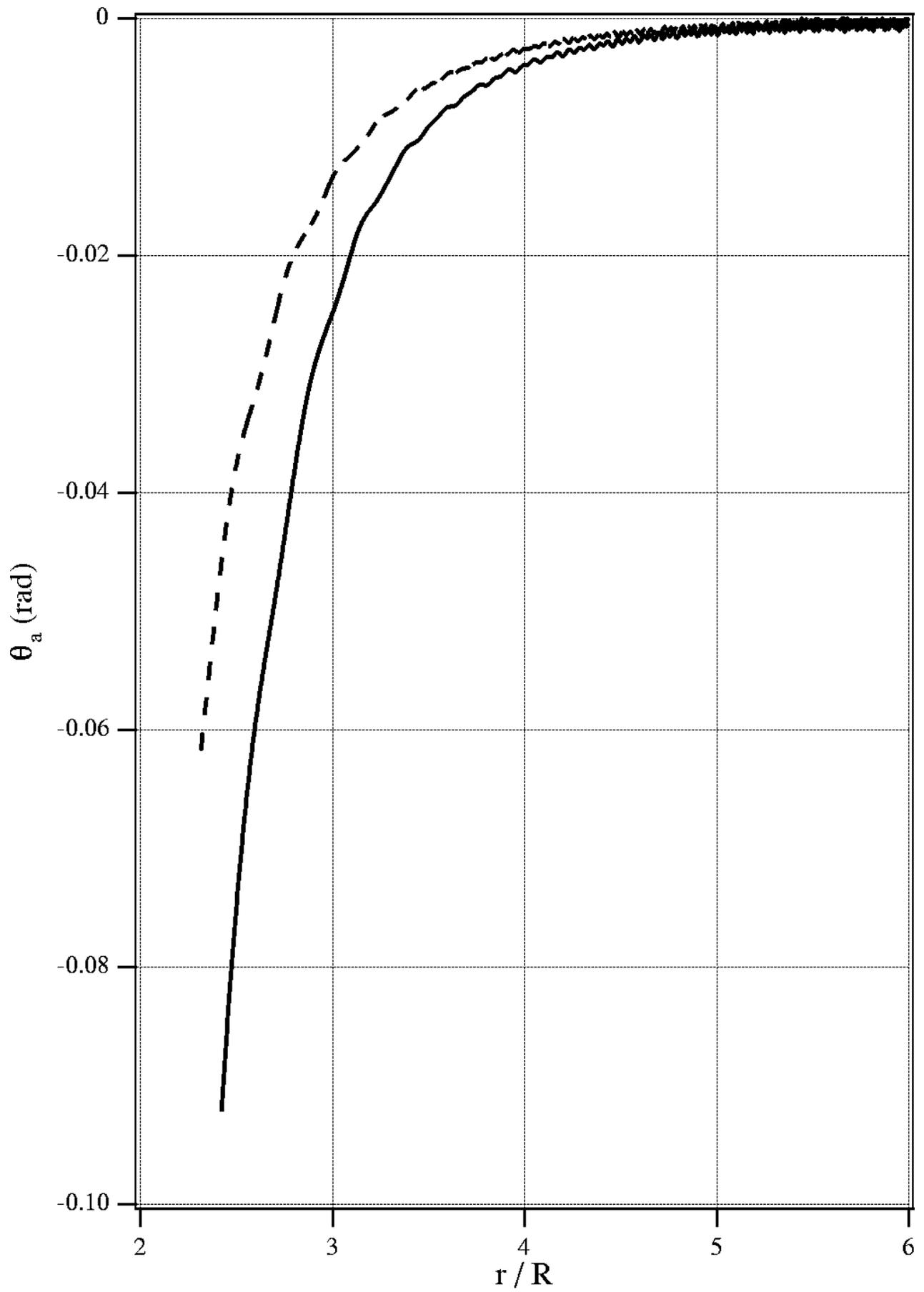
{\bf Fig.5:} 
The lag angle for the binary without the initial spin. 
The solid and dashed lines correspond
 to $n=0.5$ and $n=1$, respectively. 
\\
\\ 
{\bf Fig.6a:} 
The gravitational wave forms $h_+$ from the binary
 with polytropic index $n$. 
The solid and dashed lines correspond to
 the result of $n=0.5$ and that of $n=1$, respectively. 
Result of point masses is also given by the dotted line
 for comparison. 
\\
\\ 
{\bf Fig.6b:} 
The same as Fig.6a, but for the gravitational wave forms
 $h_\times$. 
\\ 
\\
{\bf Fig.7a:} 
The gravitational wave forms $h_+$ of the binary
 with $n=0.5$ from various initial spin states. 
The solid, dotted and dashed lines correspond to $\chi = 0, 1, 2$. 
\\ 
\\
{\bf Fig.7b:} 
The same as Fig.7a, but the initial spin is counter-rotating. 
The solid, dotted and dashed lines correspond to $\chi = 0, -1, -2$. 
\\ 
\\
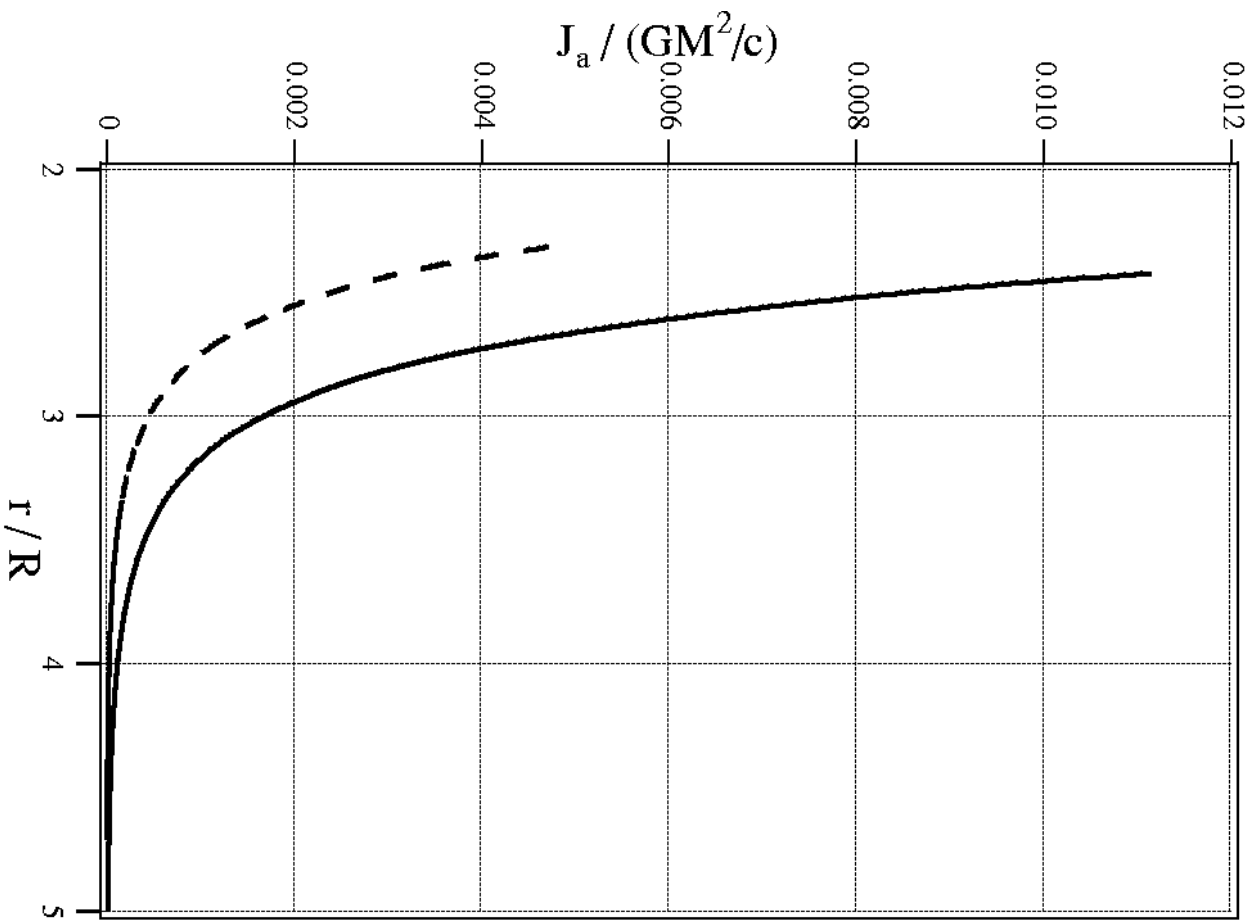
{\bf Fig.8:} 
Variation of stellar spin with all corrections. 
The solid line represents the evolution for $n=0.5$ model,
 and the dashed line for $n=1$. 
The initial spin is zero. 
\\
\\
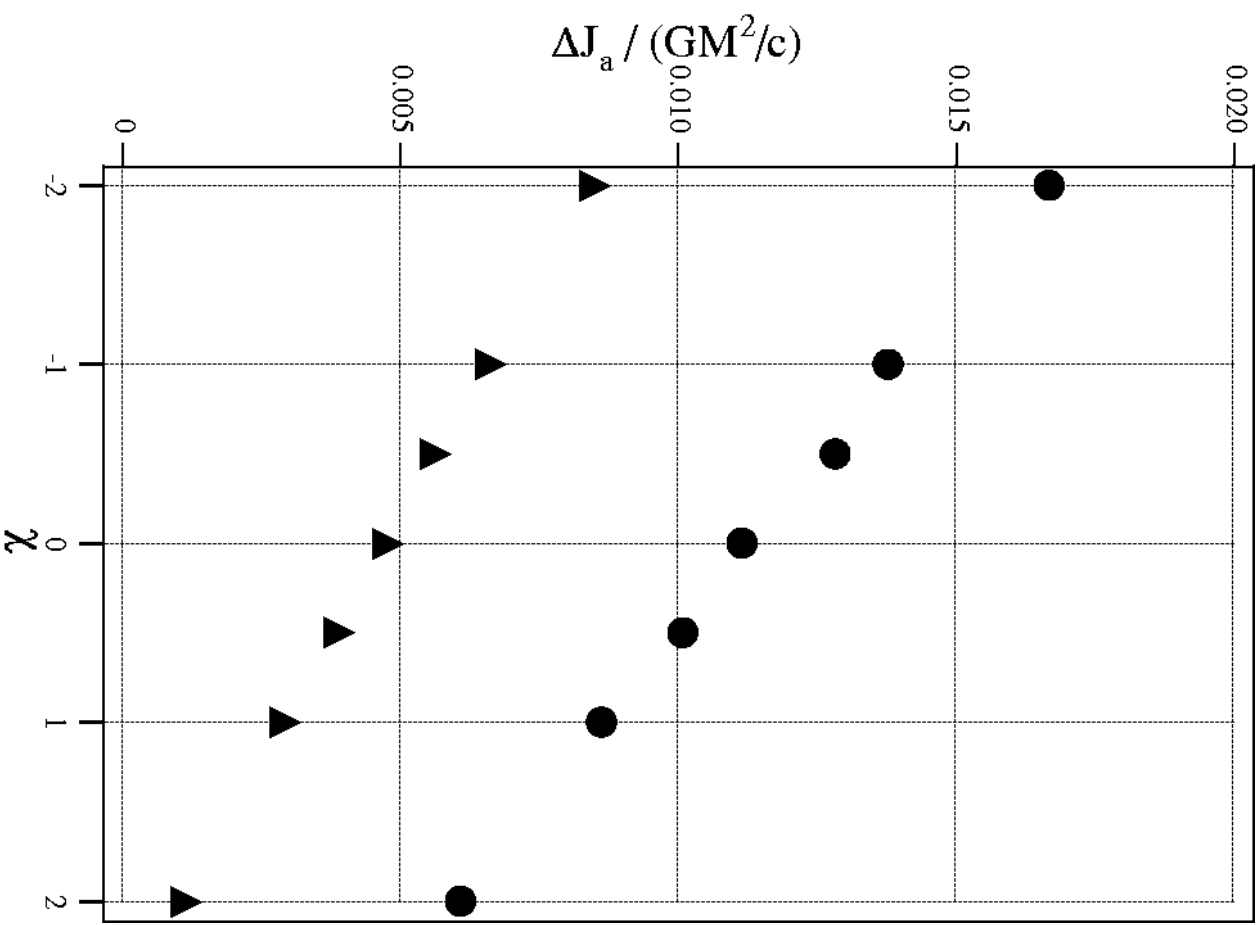
{\bf Fig.9:} 
Total change of the spin angular momentum for various
 initial $\chi$ values. 
The filled circles and triangles correspond to
 $n=0.5$ and $n=1$ model. 

%%%%%%%%%%%%%%%%%%%%
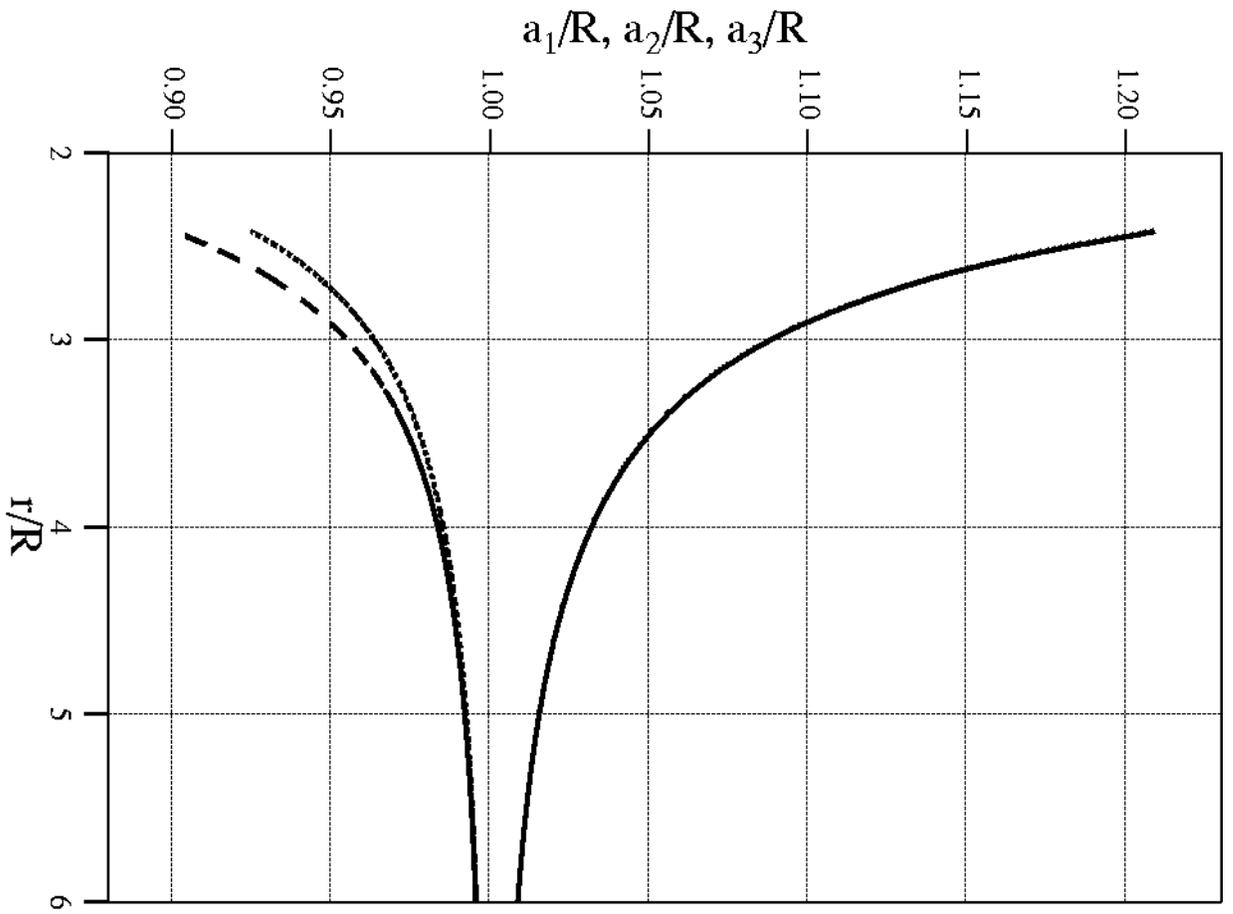
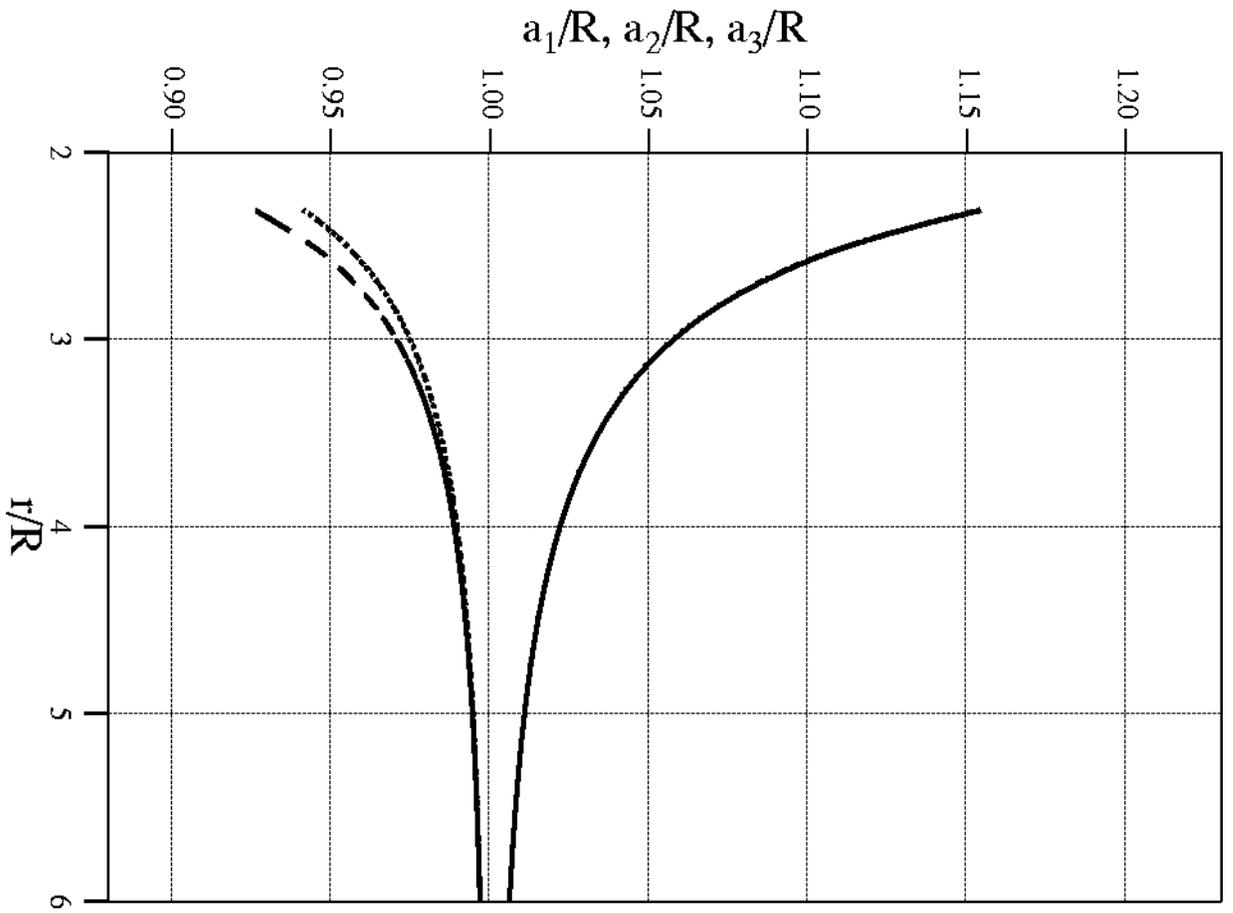
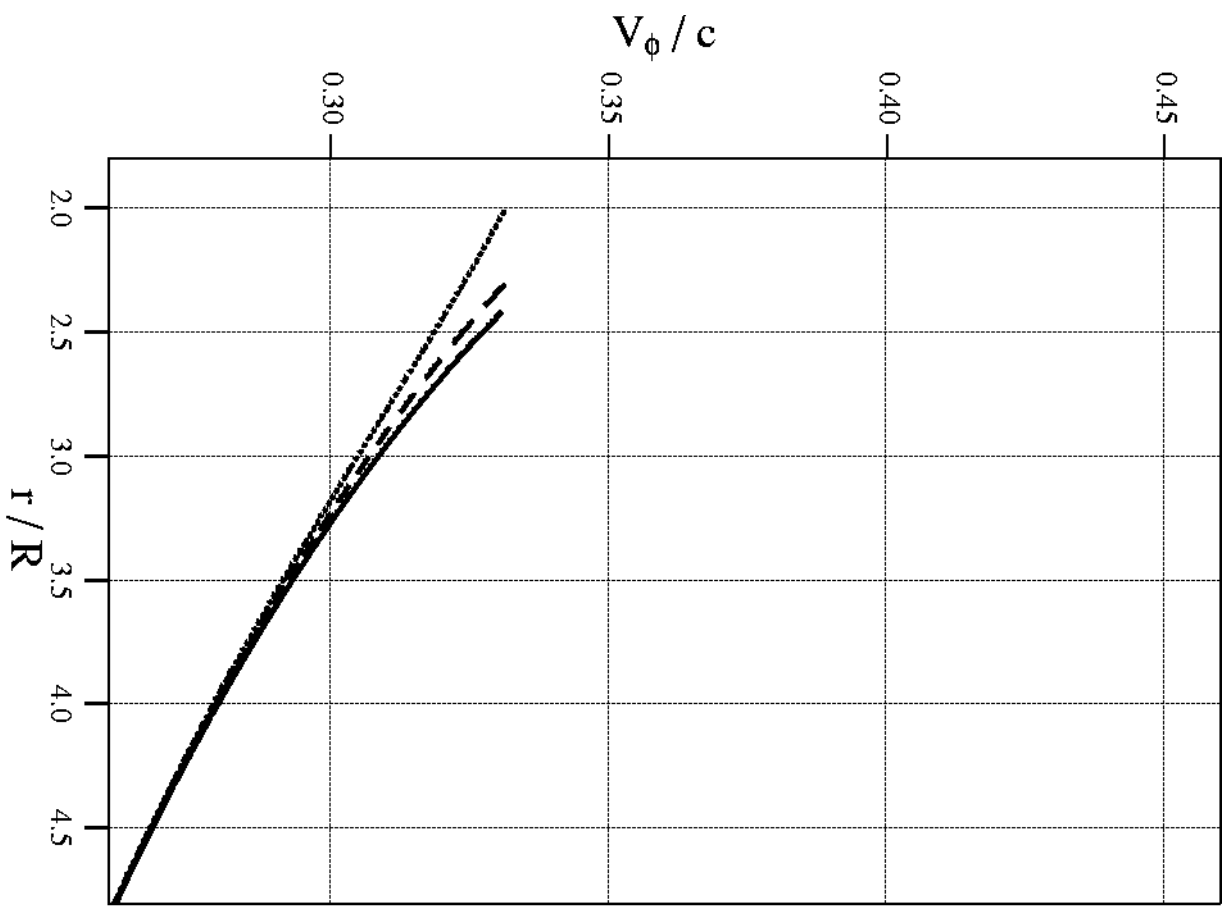
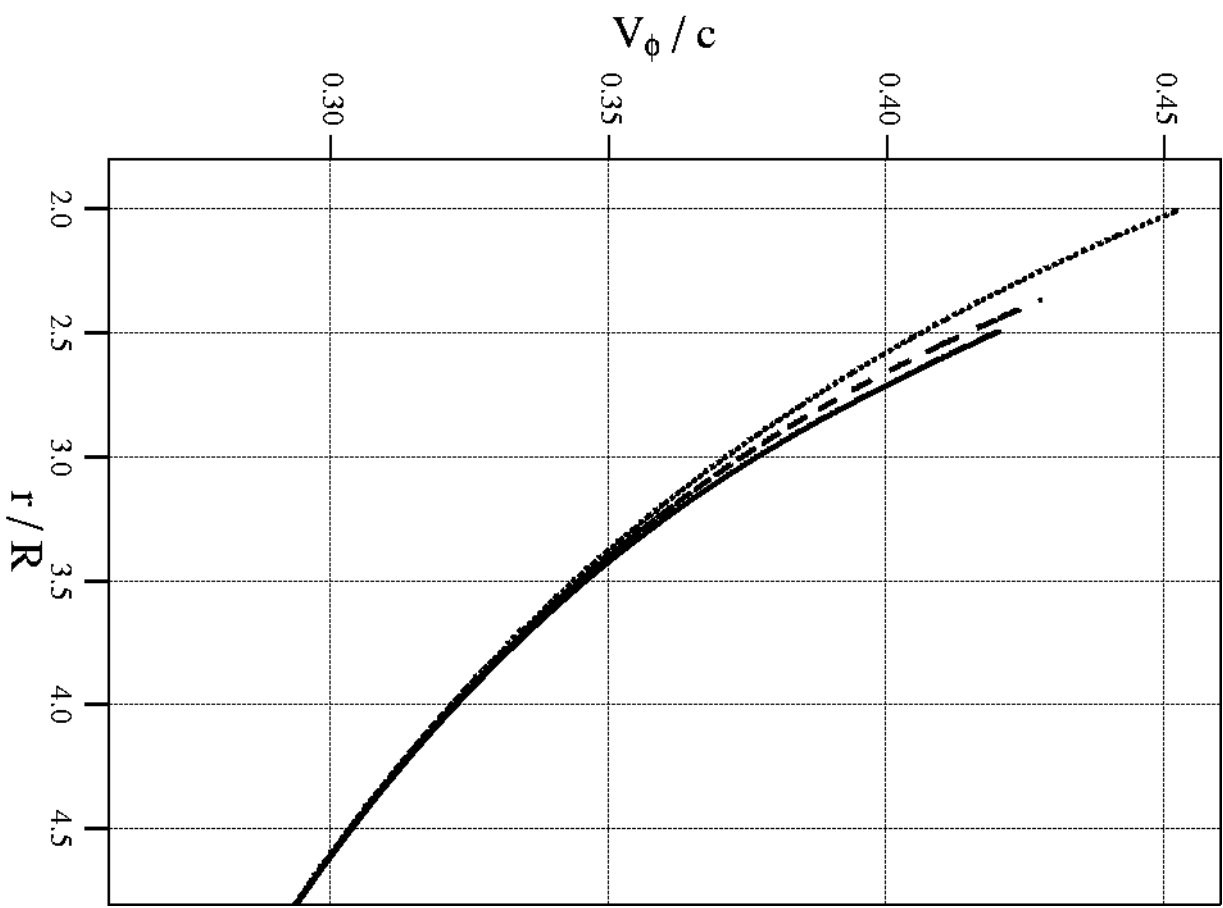
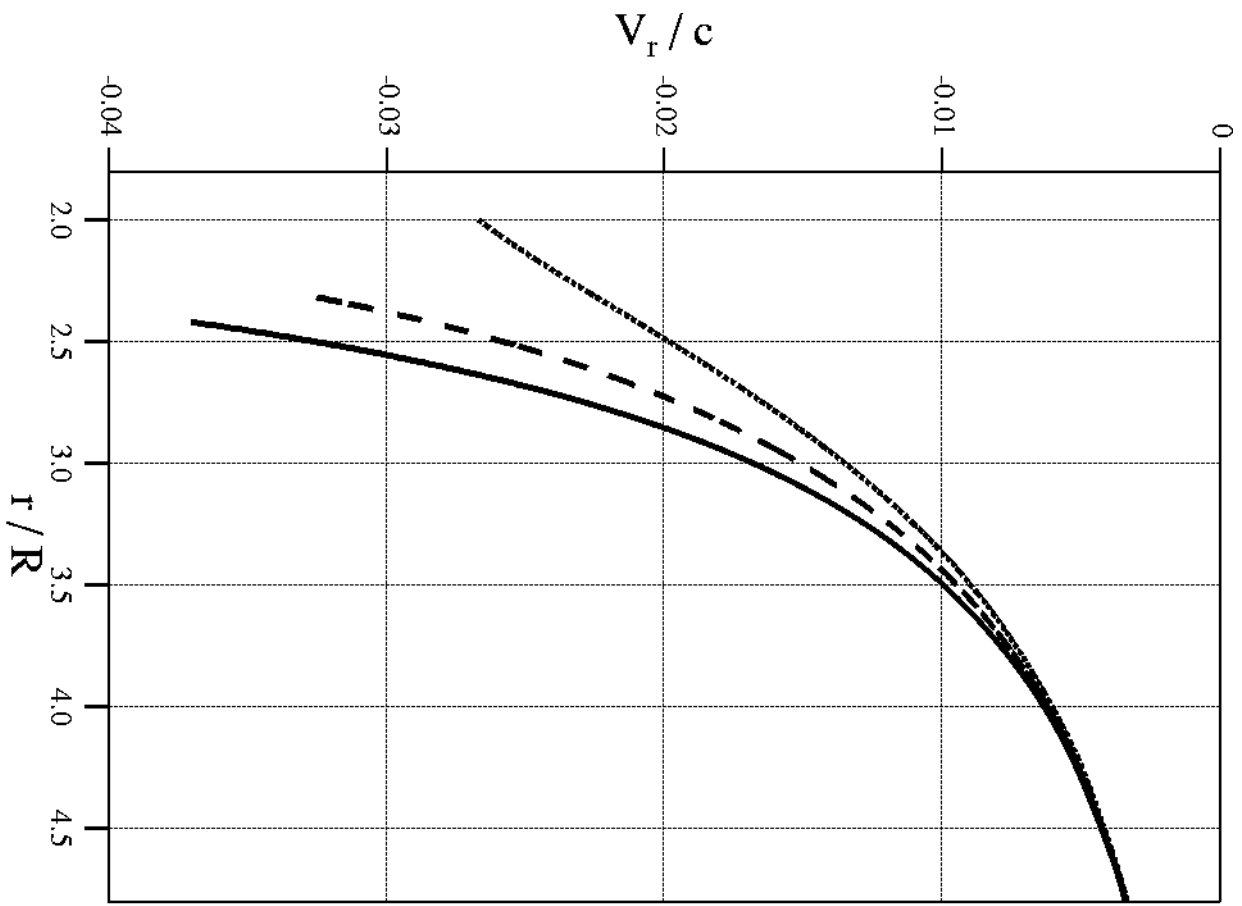
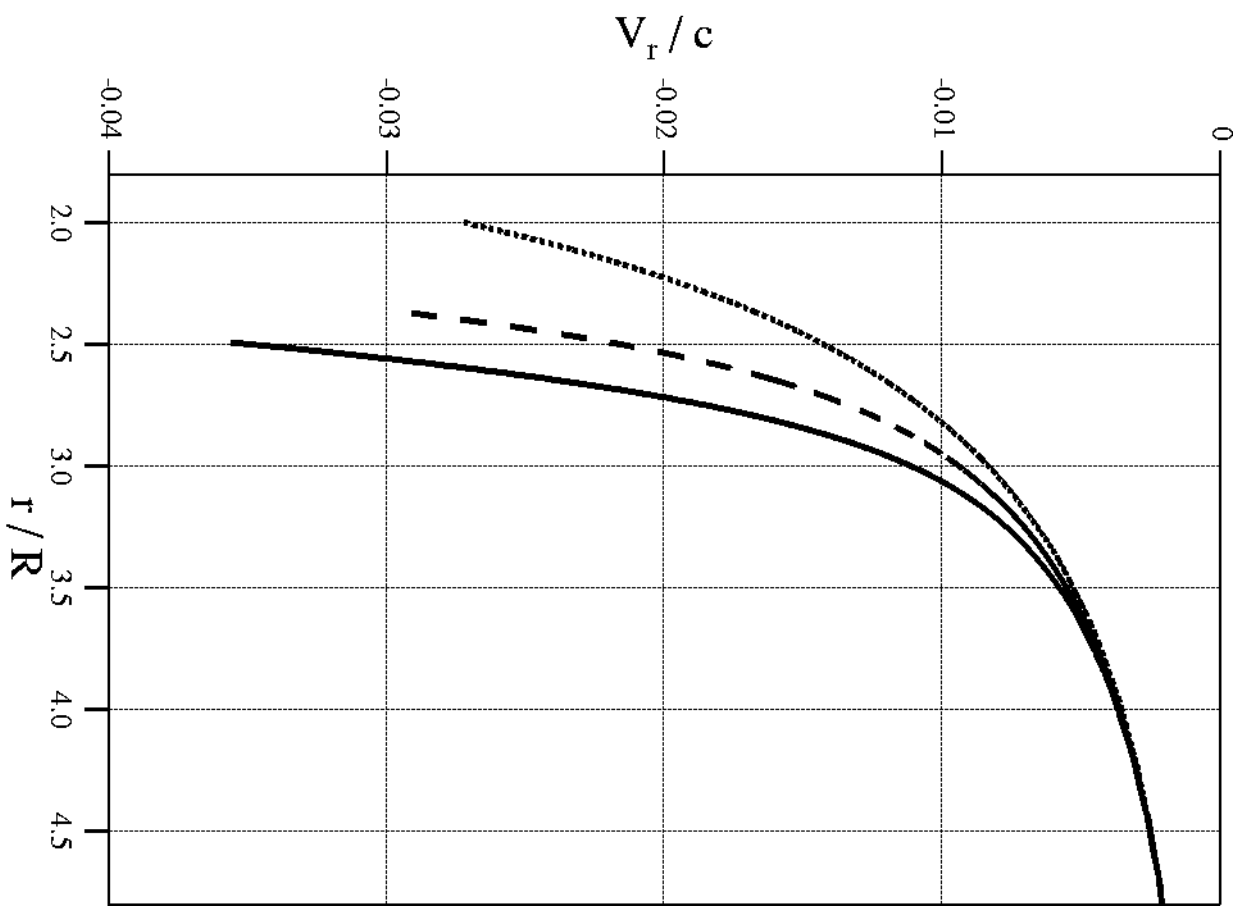
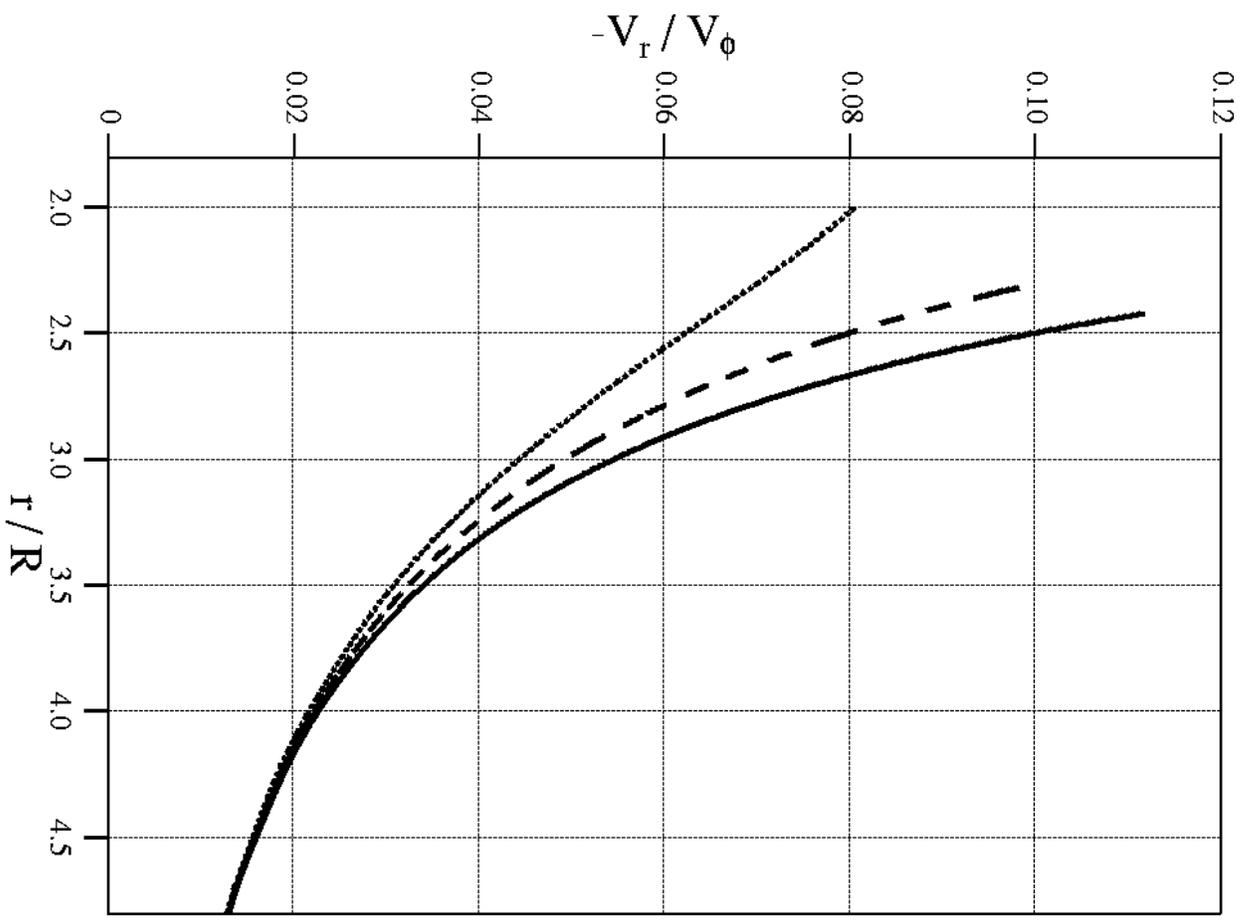
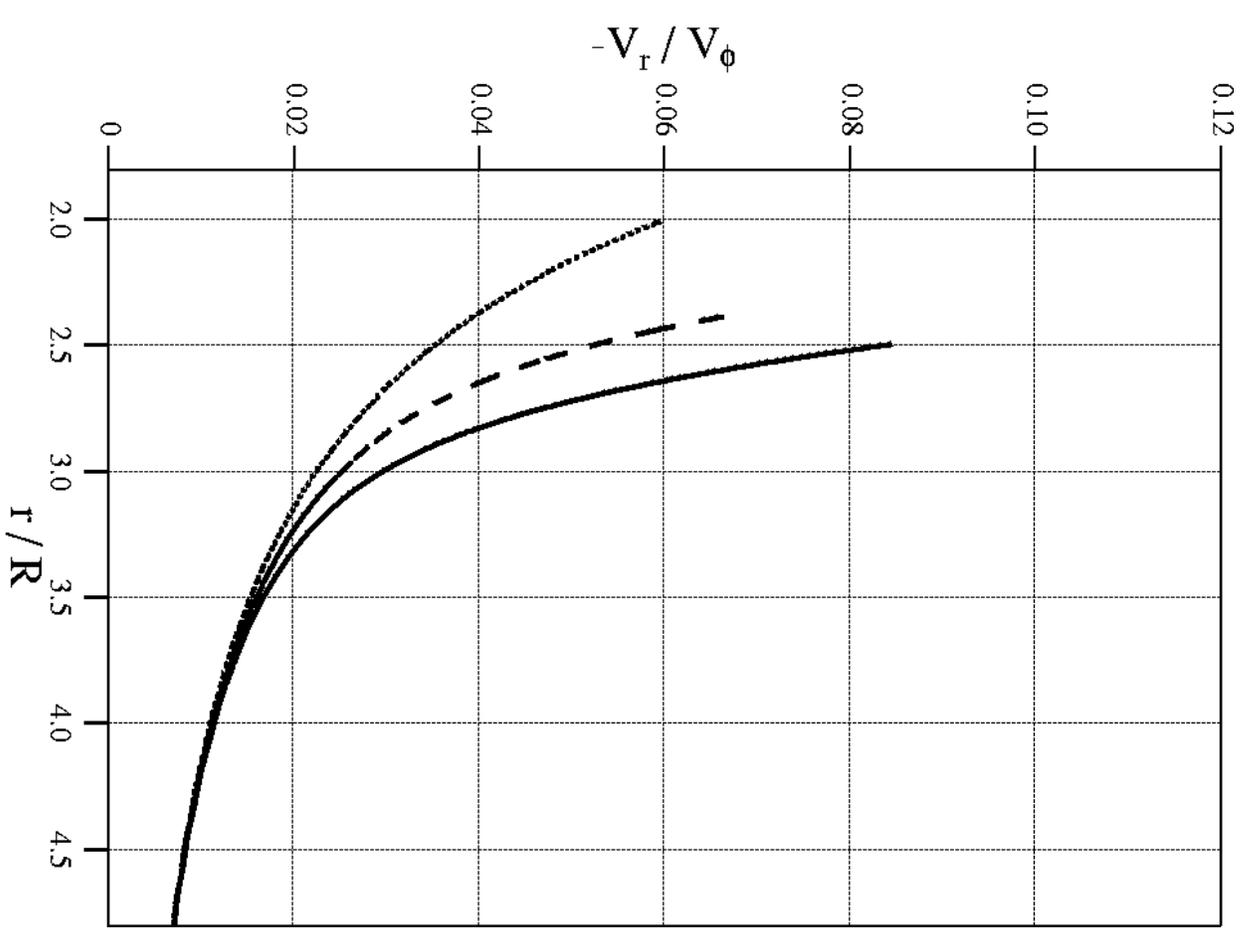
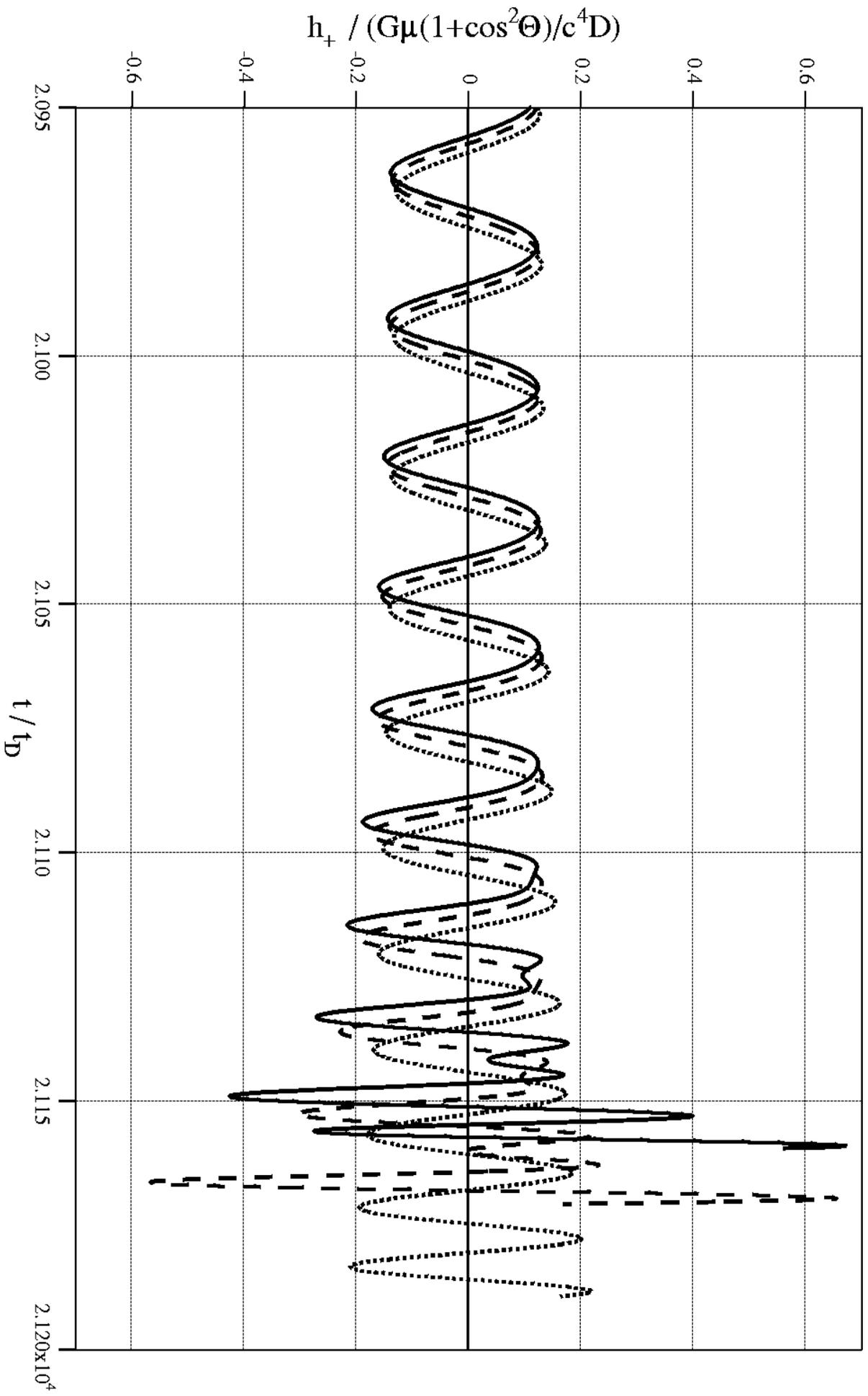
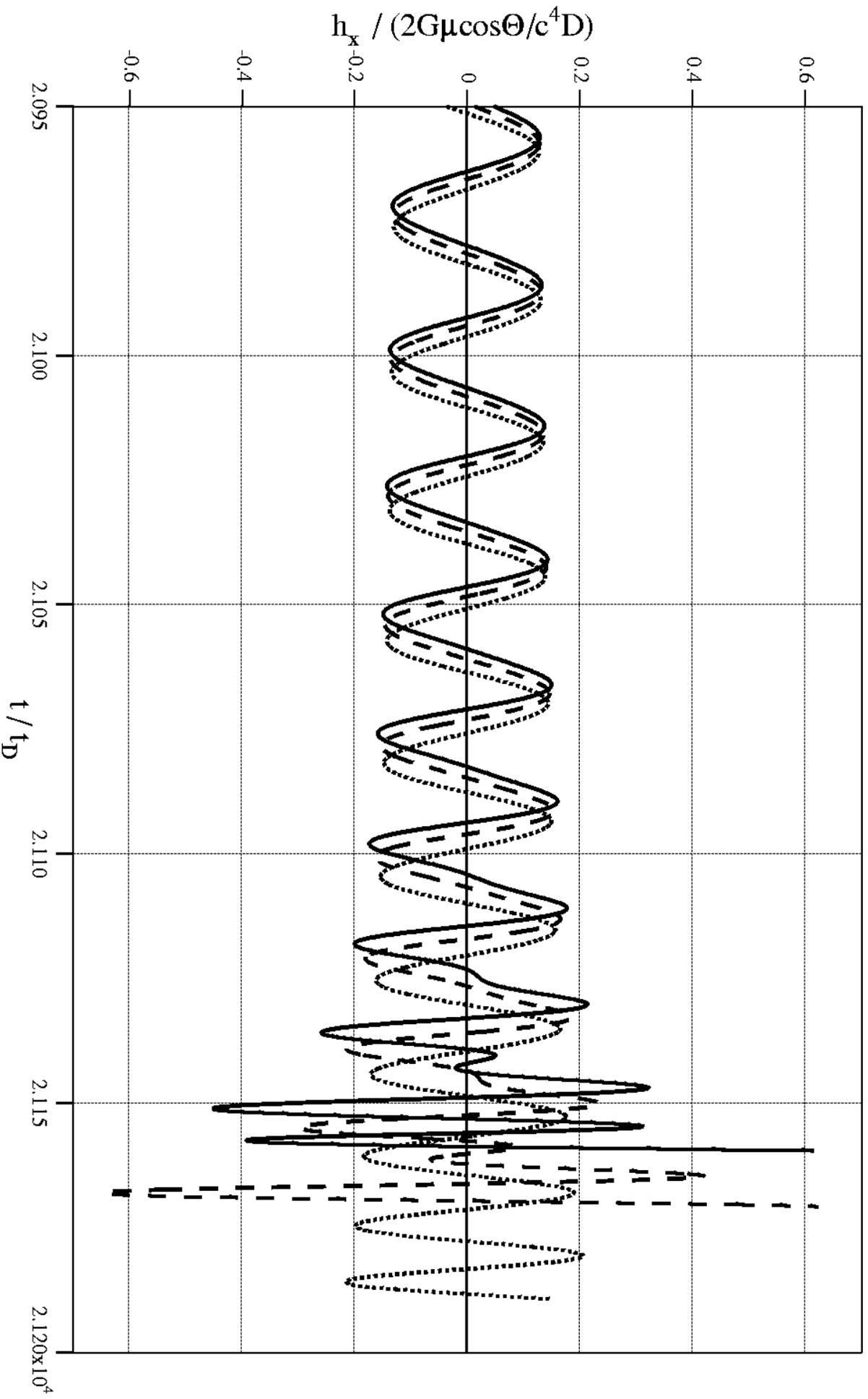
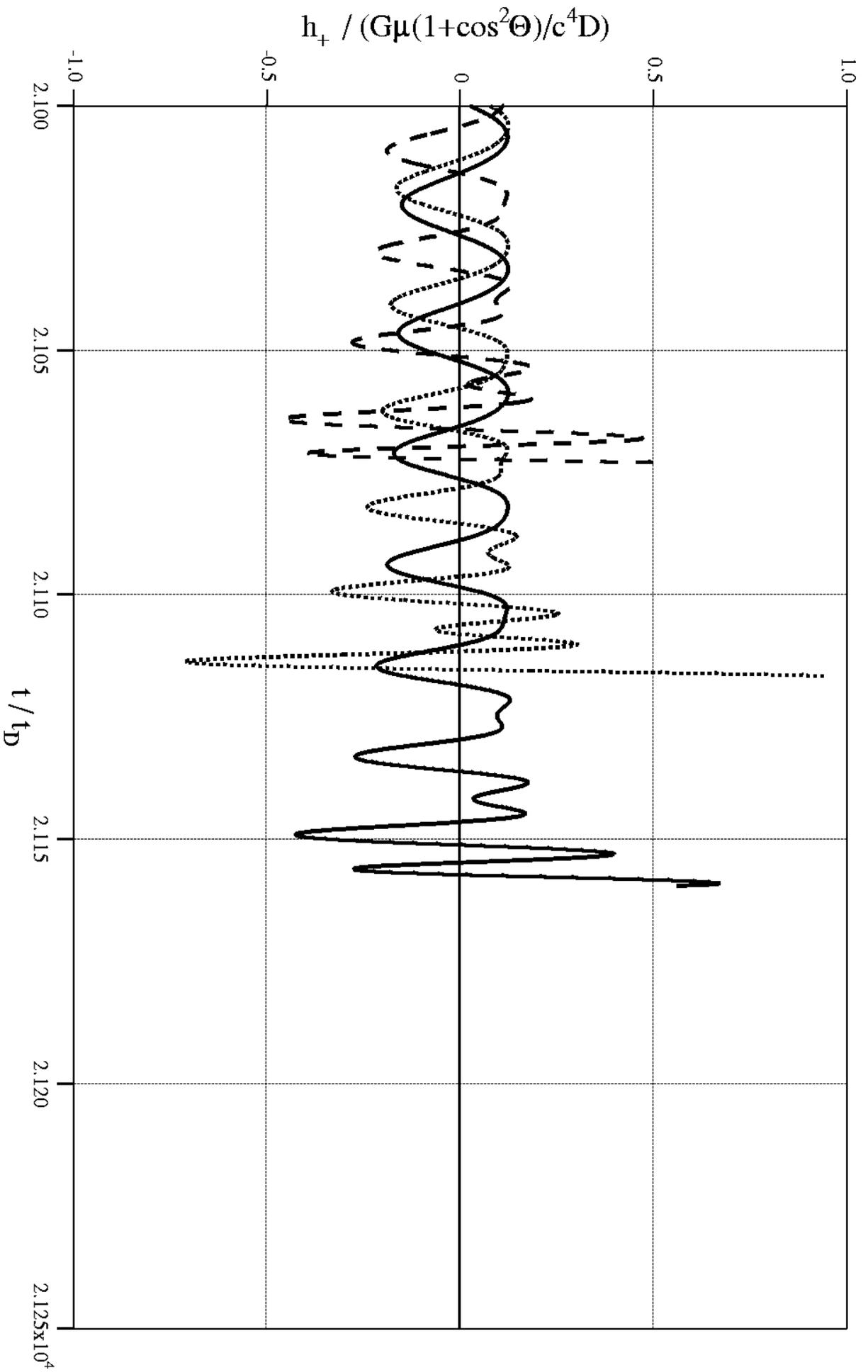
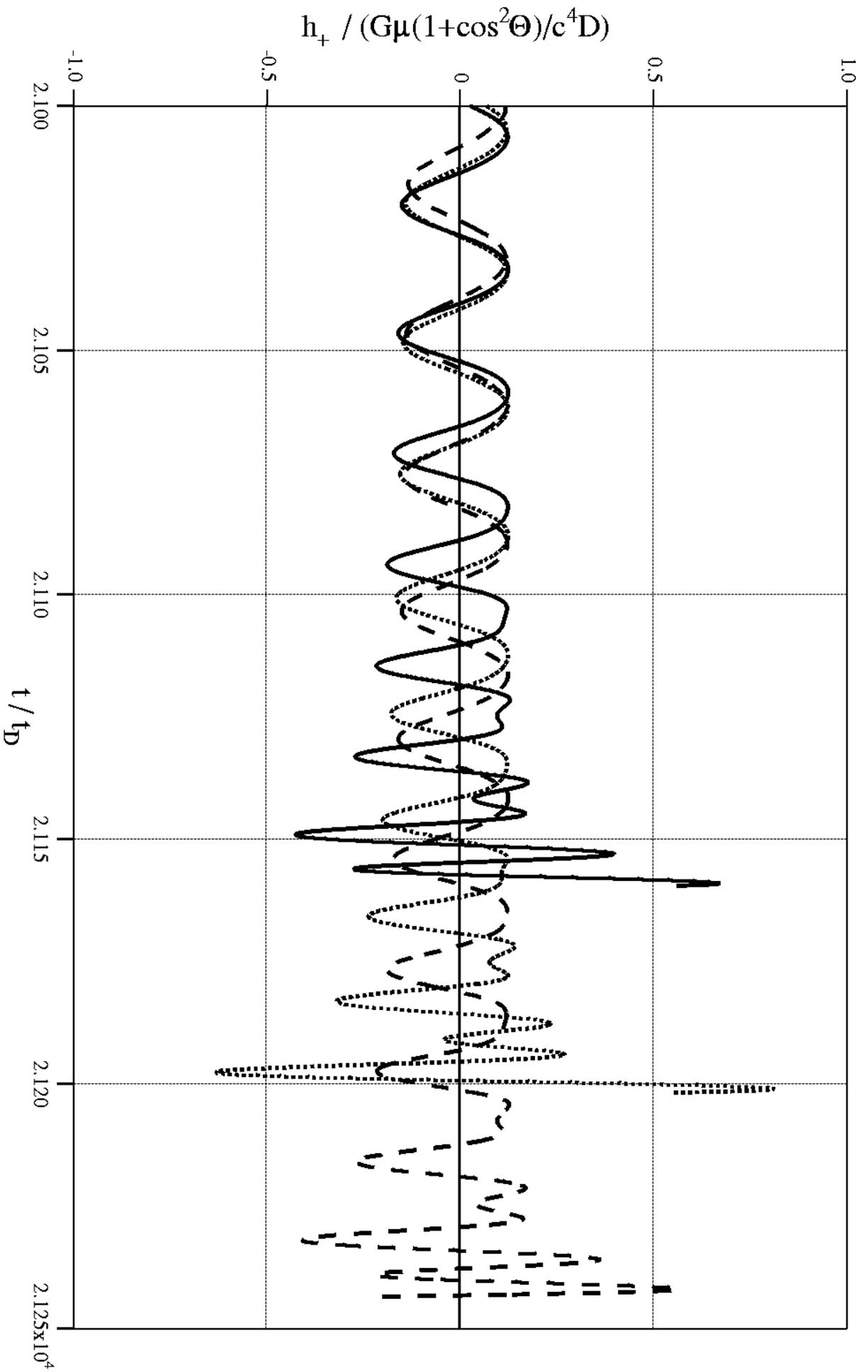
\end{document}